\documentclass[a4paper]{jpconf}
\usepackage{graphicx}
\begin{document}
\title{Early thermalization of quark-gluon matter by elastic 3-to-3 scattering}

\author{Xiao-Ming Xu}

\address{Department of Physics, Shanghai University, Baoshan, Shanghai 200444, 
China}

\ead{xmxu@mail.shu.edu.cn}

\begin{abstract}
The early thermalization is crucial to the quark-gluon plasma as a perfect
liquid and results from many-body scattering. We calculate squared amplitudes
for elastic parton-parton-parton scattering in perturbative QCD. Transport
equations with the squared amplitudes are established and solved to obtain the
thermalization time of initially produced quark-gluon matter and the initial
temperature of quark-gluon plasma. We find that the thermalization times of
quark matter and gluon matter are different.
\end{abstract}

\section{Introduction}
\medskip
Quark-gluon matter initially produced in Au-Au collisions at RHIC energies is 
not in a thermal state and has not a temperature since gluon and quark 
distributions are anisotropic in momentum space. After a period of time of
$t_{\rm therm}$, quark-gluon matter is in a thermal state and has a temperature
while the distributions are isotropic, i.e. quark-gluon matter is a
quark-gluon plasma. The transition from initially produced quark-gluon matter 
to the quark-gluon plasma is a thermalization process and $t_{\rm therm}$ is
the thermalization time. It had not been expected
that the thermalization time is small before the measured
elliptic flow coefficient of hadrons was first reported. In order to explain
the measured data, hydrodynamic calculations assume early thermalization
and ideal relativistic fluid flow \cite{kolb,teaney,hirano}. Early 
thermalization is that a thermal state is achieved with a time less than 1
fm/$c$ from the moment when quark-gluon matter is initially created in 
heavy-ion collisions. Different thermalization times have been given in
different hydrodynamic models \cite{PHENIX}. Strongly interacting matter in
thermal equilibrium can very easily flow in any direction without any breakup,
i.e. this matter is a
perfect liquid. Strong interaction of particles inside dense matter establishes
a thermal state rapidly. This means that the quark-gluon plasma formed by the
early thermalization is a perfect liquid. Therefore, the early thermalization
is crucial to the perfect liquid the quark-gluon plasma.

The early thermalization is an assumption in hydrodynamic calculations. We need
to understand the early thermalization. Then, two questions are raised. One is
how initially produced quark-gluon matter establishes a thermal state, another
is when initially produced quark-gluon matter establishes a thermal state. To 
answer the two questions, we examine the occurrence probabilities of two-parton
scattering, three-parton scattering and so on in quark-gluon matter. This is 
motivated by the thought that thermalization is related to two-body scattering
\cite{shuryak,wong,nayak,shin}. One thousand and five hundred partons are
generated from HIJING within $-0.3<z<0.3$ fm in the longitudinal direction and
$r<6.4$ fm in the transverse direction in central
Au-Au collisions at $\sqrt {s_{NN}}
=200$ GeV. This volume of partons corresponds to a parton number density of 
19.4 fm$^{-3}$. The maxima of the numbers of $n$-parton scattering with $n=2,
\cdot \cdot \cdot, 7$ are 750, 500, 375, 300, 250 and 214, respectively. The
occurrence probability can be represented by the number of $n$-parton
scattering divided by the sum of the maximum numbers of two-, three-,
$\cdot \cdot \cdot$, and seven-parton scattering. The occurrence probability
increases with increasing interaction range. At an interaction range of 0.62
fm the 2-parton scattering and 3-parton scattering have occurrence
probabilities of 0.3 and 0.2, respectively \cite{xmxu1}. Therefore, the
three-parton scattering is important due to the high parton number density.
One effect of elastic three-gluon scattering is the early thermalization of
gluon matter initially produced in central Au-Au collisions \cite{xmxu2}. The
early thermalization is an effect of many-body interaction!

Thermalization of quark-gluon matter at a high number density is governed by
the two-parton scattering and the three-parton scattering. This is the answer
to the first question. An answer to the second question is given in the next
two sections.

\section{Elastic three-parton scattering}
\medskip
Thermalization of water (air) is governed by elastic two-body
scattering of water molecules (air molecules). Thermalization by elastic
two-body scattering is conventional. Therefore, two-parton scattering is first
taken into account to examine thermalization of quark-gluon matter. Elastic
parton-parton scattering was studied by Cutler and Sivers in \cite{cutler}
and by Combridge, Kripfganz and Ranft in \cite{combridge} in
perturbative QCD. Spin- and color-averaged squared amplitudes of order 
$\alpha^2_{\rm s}$ were derived by hand and expressions for the squared
amplitudes are presented in terms of the Mandelstam variables in 
\cite{cutler,combridge}. It is very convenient to use the expressions.

However, it is impractical to derive squared amplitudes for elastic 
parton-parton-parton scattering by hand and the squared amplitudes have to be
derived from Fortran code. This is because many Feynman diagrams even at order
$\alpha^4_{\rm s}$ are involved in the elastic 3-to-3 scattering. For example,
there are 72 diagrams for elastic scattering of one gluon and two
identical quarks and 76 diagrams for elastic gluon-quark-antiquark
scattering \cite{xmxu3}. Only one Feynman diagram has two four-gluon vertices
and it is shown in figure 1. 
\begin{figure}[h]
\includegraphics[width=14pc]{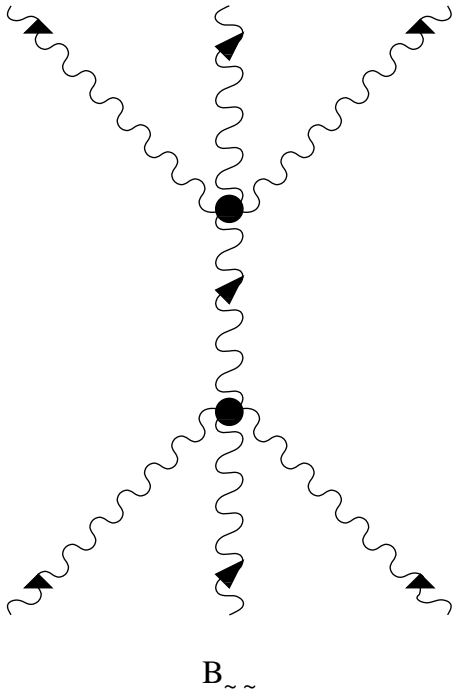}\hspace{2pc}%
\begin{minipage}[b]{20pc}\caption{\label{ggg1}Scattering of three gluons. The
wiggly lines stand for gluons.}
\end{minipage}
\end{figure}
The squared amplitude for the diagram takes the shortest expression among all
diagrams involved in elastic parton-parton-parton scattering at the tree level.
Including the average over the spin and color states of the three initial
gluons, the spin- and color-summed squared amplitude is \cite{xmxu4}
\begin{displaymath}
\frac {118098\pi^4\alpha^4_{\rm s}}{s^2}
\end{displaymath}
where $s$ is the square of the sum of the four-momenta of the three initial
gluons. The expression has only one term, but the squared amplitudes for other
diagrams have more terms with the reasons as follows. The other diagrams must
have at least one triple-gluon vertex or the trace of a product of Dirac
matrices. According to the Feynman rules the triple-gluon vertex consists of 
three terms. The trace of a product of even four Dirac matrices gives three
terms. Then, the squared amplitude for each of the other Feynman diagrams or
any interference term of different diagrams must have at least three terms. 

According to the numbers of the triple-gluon vertex and the four-gluon
vertex, Feynman diagrams are classified and their squared amplitudes are
organized. For instance, the 72 diagrams for elastic gluon-quark-quark
scattering are separated into 3 classes \cite{xmxu3}. 
The first, second and third classes contain 40
diagrams with no triple-gluon vertex and no four-gluon vertex,
24 diagrams with one triple-gluon vertex and no four-gluon vertex
and 8 diagrams with two triple-gluon vertices or one four-gluon vertex,
respectively. The 76 diagrams for elastic gluon-quark-antiquark scattering are
also divided into 3 classes even though quark-antiquark annihilation may occur
\cite{xmxu3}. The first class contains 40
diagrams with no triple-gluon vertex and no four-gluon vertex: 
20 diagrams with no quark-antiquark
annihilation and 20 diagrams with the annihilation. The second class contains
28 diagrams with one triple-gluon vertex and no four-gluon vertex: 12 with no
quark-antiquark annihilation and 16 with the annihilation. The third class 
contains 8 diagrams with two triple-gluon vertices or one four-gluon vertex: 
4 with no quark-antiquark annihilation and 4 with the annihilation.

\section{Thermalization}
\medskip
We establish transport equations for quark-gluon matter which consists of
gluons, quarks and antiquarks with up and down flavors. We assume that the
quark distribution is symmetric in flavor and is identical with the antiquark
distribution. The quark and gluon distribution functions are denoted by 
$f_{qi}$ and $f_{{\rm g}i}$, respectively,
where $i$ labels the $i$th quark or gluon. With the elastic
parton-parton scattering and the elastic parton-parton-parton scattering, the
transport equation for gluons is \cite{xmxu3}
\begin{eqnarray}
\frac {\partial f_{{\rm g}1}}{\partial t}
& + & \vec {\rm v}_1 \cdot \vec {\nabla}_{\vec {r}} f_{{\rm g}1} 
~~ = ~~ -\frac {1}{2E_1} \int \frac {d^3p_2}{(2\pi)^32E_2}
\frac {d^3p_3}{(2\pi)^32E_3} \frac {d^3p_4}{(2\pi)^32E_4}
(2\pi)^4 \delta^4(p_1+p_2-p_3-p_4)
         \nonumber    \\
& &
\times \left\{ \frac {{\rm g}_G}{2} 
\mid {\cal M}_{{\rm g}{\rm g} \to {\rm g}{\rm g}} \mid^2
[f_{{\rm g}1}f_{{\rm g}2}(1+f_{{\rm g}3})(1+f_{{\rm g}4})
-f_{{\rm g}3}f_{{\rm g}4}(1+f_{{\rm g}1})(1+f_{{\rm g}2})]  \right.
         \nonumber    \\
& &
+ {\rm g}_Q ( \mid {\cal M}_{{\rm g}u \to {\rm g}u} \mid^2
+ \mid {\cal M}_{{\rm g}d \to {\rm g}d} \mid^2
+ \mid {\cal M}_{{\rm g}\bar {u} \to {\rm g}\bar {u}} \mid^2
+ \mid {\cal M}_{{\rm g}\bar {d} \to {\rm g}\bar {d}} \mid^2 )
         \nonumber    \\
& &
\left.
\times [f_{{\rm g}1}f_{q2}(1+f_{{\rm g}3})(1-f_{q4})
-f_{{\rm g}3}f_{q4}(1+f_{{\rm g}1})(1-f_{q2})]
    \right\}
         \nonumber    \\
& &
-\frac {1}{2E_1} \int \frac {d^3p_2}{(2\pi)^32E_2}
\frac {d^3p_3}{(2\pi)^32E_3} \frac {d^3p_4}{(2\pi)^32E_4}
\frac {d^3p_5}{(2\pi)^32E_5} \frac {d^3p_6}{(2\pi)^32E_6}
         \nonumber    \\
& &
\times (2\pi)^4 \delta^4(p_1+p_2+p_3-p_4-p_5-p_6) 
\left\{ \frac {{\rm g}_G^2}{12} 
\mid {\cal M}_{{\rm g}{\rm g}{\rm g} \to {\rm  g}{\rm g}{\rm g}} \mid^2 \right.
         \nonumber    \\
& &
\times [f_{{\rm g}1}f_{{\rm g}2}f_{{\rm g}3}
(1+f_{{\rm g}4})(1+f_{{\rm g}5})(1+f_{{\rm g}6})-
f_{{\rm g}4}f_{{\rm g}5}f_{{\rm g}6}
(1+f_{{\rm g}1})(1+f_{{\rm g}2})(1+f_{{\rm g}3})]
         \nonumber    \\
& &
+ \frac {{\rm g}_G{\rm g}_Q}{2} 
( \mid {\cal M}_{{\rm g}{\rm g}u \to {\rm g}{\rm g}u} \mid^2
+\mid {\cal M}_{{\rm g}{\rm g}d \to {\rm g}{\rm g}d} \mid^2
+\mid {\cal M}_{{\rm g}{\rm g}\bar {u} \to {\rm g}{\rm g}\bar {u}} \mid^2 
+\mid {\cal M}_{{\rm g}{\rm g}\bar {d} \to {\rm g}{\rm g}\bar {d}} \mid^2 )
         \nonumber    \\
& &
\times [f_{{\rm g}1}f_{{\rm g}2}f_{q3}
(1+f_{{\rm g}4})(1+f_{{\rm g}5})(1-f_{q6})
-f_{{\rm g}4}f_{{\rm g}5}f_{q6}
(1+f_{{\rm g}1})(1+f_{{\rm g}2})(1-f_{q3})]
         \nonumber    \\
& &
+ {\rm g}_Q^2 [\frac {1}{4} \mid {\cal M}_{{\rm g}uu \to {\rm g}uu} \mid^2
+\frac {1}{2} ( \mid {\cal M}_{{\rm g}ud \to {\rm g}ud} \mid^2
              + \mid {\cal M}_{{\rm g}du \to {\rm g}du} \mid^2 )
+\frac {1}{4} \mid {\cal M}_{{\rm g}dd \to {\rm g}dd} \mid^2
         \nonumber    \\
& &
+ \mid {\cal M}_{{\rm g}u\bar {u} \to {\rm g}u\bar {u}} \mid^2
    + \mid {\cal M}_{{\rm g}u\bar {d} \to {\rm g}u\bar {d}} \mid^2
    + \mid {\cal M}_{{\rm g}d\bar {u} \to {\rm g}d\bar {u}} \mid^2
    + \mid {\cal M}_{{\rm g}d\bar {d} \to {\rm g}d\bar {d}} \mid^2
         \nonumber         \\
& &
+\frac {1}{4} \mid {\cal M}_{{\rm g}\bar {u}\bar {u}
                             \to {\rm g}\bar {u}\bar {u}} \mid^2
    +\frac {1}{2} ( \mid {\cal M}_{{\rm g}\bar {u}\bar {d}
                             \to {\rm g}\bar {u}\bar {d}} \mid^2 
    + \mid {\cal M}_{{\rm g}\bar {d}\bar {u} 
                             \to {\rm g}\bar {d}\bar {u}} \mid^2 )
+ \frac {1}{4} \mid {\cal M}_{{\rm g}\bar {d}\bar {d} 
                             \to {\rm g}\bar {d}\bar {d}} \mid^2 ]
         \nonumber    \\
& &
\left. \times [f_{{\rm g}1}f_{q2}f_{q3}(1+f_{{\rm g}4})(1-f_{q5})(1-f_{q6})
     -f_{{\rm g}4}f_{q5}f_{q6}(1+f_{{\rm g}1})(1-f_{q2})(1-f_{q3})] \right\} ,
         \nonumber    \\
\end{eqnarray}
and the transport equation for up-quarks is
\begin{eqnarray}
\frac {\partial f_{q1}}{\partial t}
& + & \vec {\rm v}_1 \cdot \vec {\nabla}_{\vec {r}} f_{q1}
~~ = ~~ -\frac {1}{2E_1} \int \frac {d^3p_2}{(2\pi)^32E_2}
\frac {d^3p_3}{(2\pi)^32E_3} \frac {d^3p_4}{(2\pi)^32E_4}
(2\pi)^4 \delta^4(p_1+p_2-p_3-p_4)
         \nonumber    \\
& &
\times \left\{ {\rm g}_G \mid {\cal M}_{u{\rm g} \to u{\rm g}} \mid^2
[f_{q1}f_{{\rm g}2}(1-f_{q3})(1+f_{{\rm g}4})
-f_{q3}f_{{\rm g}4}(1-f_{q1})(1+f_{{\rm g}2})]  \right.
         \nonumber    \\
& &
+ {\rm g}_Q (\frac {1}{2} \mid {\cal M}_{uu \to uu} \mid^2
+ \mid {\cal M}_{ud \to ud} \mid^2 
+ \mid {\cal M}_{u\bar {u} \to u\bar {u}} \mid^2
+ \mid {\cal M}_{u\bar {d} \to u\bar {d}} \mid^2 )
         \nonumber    \\
& &
\times \left. [f_{q1}f_{q2}(1-f_{q3})(1-f_{q4})
-f_{q3}f_{q4}(1-f_{q1})(1-f_{q2})]   \right\}
         \nonumber    \\
& &
-\frac {1}{2E_1} \int \frac {d^3p_2}{(2\pi)^32E_2}
\frac {d^3p_3}{(2\pi)^32E_3} \frac {d^3p_4}{(2\pi)^32E_4}
\frac {d^3p_5}{(2\pi)^32E_5} \frac {d^3p_6}{(2\pi)^32E_6}
         \nonumber    \\
& &
\times (2\pi)^4 \delta^4(p_1+p_2+p_3-p_4-p_5-p_6)
\left\{ \frac {{\rm g}_G^2}{4} 
\mid {\cal M}_{u{\rm g}{\rm g} \to u{\rm g}{\rm g}} \mid^2  \right.
         \nonumber    \\
& &
\times [f_{q1}f_{{\rm g}2}f_{{\rm g}3}(1-f_{q4})
(1+f_{{\rm g}5})(1+f_{{\rm g}6})
-f_{q4}f_{{\rm g}5}f_{{\rm g}6}(1-f_{q1})(1+f_{{\rm g}2})(1+f_{{\rm g}3})]
          \nonumber     \\
& &
+{\rm g}_Q{\rm g}_G ( \frac {1}{2} 
\mid {\cal M}_{uu{\rm g} \to uu{\rm g}} \mid^2
+\mid {\cal M}_{ud{\rm g} \to ud{\rm g}} \mid^2
+\mid {\cal M}_{u\bar {u}{\rm g} \to u\bar {u}{\rm g}} \mid^2
+\mid {\cal M}_{u\bar {d}{\rm g} \to u\bar {d}{\rm g}} \mid^2 )
         \nonumber     \\
& &
\times [f_{q1}f_{q2}f_{{\rm g}3}(1-f_{q4})(1-f_{q5})(1+f_{{\rm g}6})
-f_{q4}f_{q5}f_{{\rm g}6}(1-f_{q1})(1-f_{q2})(1+f_{{\rm g}3})]
          \nonumber     \\
& &
+ {\rm g}_Q^2 [\frac {1}{12} \mid {\cal M}_{uuu \to uuu} \mid^2
+\frac {1}{4} ( \mid {\cal M}_{uud \to uud} \mid^2
              + \mid {\cal M}_{udu \to udu} \mid^2 )
+\frac {1}{4} \mid {\cal M}_{udd \to udd} \mid^2
         \nonumber    \\
& &
+\frac {1}{2} \mid {\cal M}_{uu\bar {u} \to uu\bar {u}} \mid^2
    +\frac {1}{2} \mid {\cal M}_{uu\bar {d} \to uu\bar {d}} \mid^2
            + \mid {\cal M}_{ud\bar {u} \to ud\bar {u}} \mid^2
            + \mid {\cal M}_{ud\bar {d} \to ud\bar {d}} \mid^2
         \nonumber         \\
& &
+\frac {1}{4} \mid {\cal M}_{u\bar {u}\bar {u}
                             \to u\bar {u}\bar {u}} \mid^2
    +\frac {1}{2} ( \mid {\cal M}_{u\bar {u}\bar {d}
                               \to u\bar {u}\bar {d}} \mid^2
+ \mid {\cal M}_{u\bar {d}\bar {u} \to u\bar {d}\bar {u}} \mid^2 )
+ \frac {1}{4}
  \mid {\cal M}_{u\bar {d}\bar {d} \to u\bar {d}\bar {d}} \mid^2 ]
         \nonumber    \\
& &
\times \left. [f_{q1}f_{q2}f_{q3}(1-f_{q4})(1-f_{q5})(1-f_{q6})
-f_{q4}f_{q5}f_{q6}(1-f_{q1})(1-f_{q2})(1-f_{q3})] \right\} ,
         \nonumber    \\
\end{eqnarray}
where $\rm \vec {v}_1$ is the parton velocity; ${\rm g}_{\rm G}$ 
and ${\rm g}_{\rm Q}$ are
the color-spin degeneracy factors; $p_i (i=1,\cdot \cdot \cdot,6)$ denote the
four-momenta of initial and final partons; $E_i$ is the energy component of
$p_i$. $\mid {\cal M}_{ab \to a^\prime 
b^\prime} \mid^2$ and $\mid {\cal M}_{abc \to a^\prime b^\prime c^\prime} 
\mid^2$ are the squared amplitudes for $a+b \to a^\prime + b^\prime$ and $a+b+c
\to a^\prime + b^\prime + c^\prime$, respectively. The squared amplitude
$\mid {\cal M}_{{\rm g}{\rm g}{\rm g} \to {\rm g}{\rm g}{\rm g}} \mid^2$ 
was obtained in \cite{xmxu2};
$\mid {\cal M}_{qqq \to qqq} \mid^2$ in \cite{xmxu5};
$\mid {\cal M}_{{\rm g}qq \to {\rm g}qq} \mid^2$,
$\mid {\cal M}_{{\rm g}q\bar {q} \to {\rm g}q\bar q} \mid^2$ and
$\mid {\cal M}_{{\rm g}\bar {q}\bar {q} \to {\rm g}\bar {q}\bar {q}} \mid^2$
in \cite{xmxu3}. $\mid {\cal M}_{{\rm g}{\rm g}q \to {\rm g}{\rm g}q} \mid^2$
and $\mid {\cal M}_{{\rm g}{\rm g}\bar {q} \to {\rm g}{\rm g}\bar q} \mid^2$
have not been obtained and are temporarily
set as zero. Similar equations for down-quarks, up-antiquarks and
down-antiquarks can be established.

Parton-parton scattering affects only two partons' momenta while
parton-parton-parton
scattering changes three partons' momenta. If the 3-to-3 scattering has
the same occurrence probability as the 2-to-2 scattering, the 3-to-3
scattering changes partons' momenta faster than the 2-to-2 scattering. Frequent
changes of momenta establish a thermal state. Indeed, the elastic 3-to-3
scattering drives anisotropic matter described by the transport equations
towards global thermal equilibrium \cite{xmxu4}, which is similar to the 
$H$-theorem. Solutions of the transport equations will show that a thermal
state is established from a moment.

In the region $-0.3<z<0.3$ fm in the longitudinal direction and $r<6.4$ fm in
the transverse direction, 1500 gluons and 1000 fermions including up quarks, 
down quarks, up antiquarks and down antiquarks are generated from HIJING for
central Au-Au collisions at $\sqrt {s_{NN}}=200$ GeV. These gluons, quarks and 
antiquarks at $t=0.2$ fm/$c$ form initially produced quark-gluon matter
and are anisotropically distributed in momentum space. Solving the transport
equations, we found that at $t=0.68$ fm/$c$ gluon matter has an isotropic
momentum distribution function, a thermal state with
the temperature $T=0.5$ GeV and a thermalization time of the order of 0.48
fm/$c$ \cite{xmxu3}. A thermal state of quark matter with $T=0.3$ GeV is
established at a later time of 1.56 fm/$c$. In other words, the thermalization
time 1.36 fm/$c$ of quark matter is longer than gluon matter \cite{xmxu3}. 
The elastic scattering of ${\rm g}q$, ${\rm g}\bar q$, ${\rm g}qq$, 
${\rm g}q\bar q$ and ${\rm g}\bar {q}\bar {q}$ gives contributions to gluon 
matter and quark matter nearly identical in thermalization. 
The difference of the thermalization times of gluon matter and
quark matter mainly comes from the difference that elastic 
${\rm g}{\rm g}{\rm g}$ (${\rm g}{\rm g}$)
scattering has a larger squared amplitude than elastic $qqq$ or $qq\bar q$
($qq$ or $q\bar q$) scattering and gluon matter is denser than quark matter.

\section{Conclusions}
\medskip
Early thermalization is an intrinsic property of initially produced quark-gluon
matter at the high number density and is an effect of many-body scattering. The
elastic 3-to-3 scattering is important at such a number density. The elastic
gluon-gluon scattering and the elastic gluon-gluon-gluon scattering lead to the
early thermalization of gluon matter. The thermalization time of quark
matter differs from that of gluon matter and we need to include the elastic
scattering of both ${\rm g}{\rm g}q$ and ${\rm g}{\rm g}\bar q$ in future work
to explore whether quark matter thermalizes rapidly or not.

\ack
This work was supported by the National Natural Science Foundation of China
under grant no 11175111.

\end{document}